_______________________________________________________________

# Speech watermarking: an approach for the forensic analysis of digital telephonic recordings

Marcos Faundez-Zanuy[a,*], Jose Juan Lucena-Molina[b], Martin Hagmüller[c]

[a]Escola Universitària Politècnica de Mataró 08303 Spain
[b]Dirección General de la Guardia Civil 28003 Spain
[c]Graz University of Technology Graz 8010 Austria

[*]Corrresponding author: Tel: +34 93 757 44 04 fax: +34 93 757 05 24
e-mail address: faundez@eupmt.es

**Abstract**
In this article, the authors discuss the problem of forensic authentication of digital audio recordings. Although forensic audio has been addressed in several articles, the existing approaches are focused on analog magnetic recordings, which are less prevalent because of the large amount of digital recorders available on the market (optical, solid state, hard disks, etc.). An approach based on digital signal processing that consists of spread spectrum techniques for speech watermarking is presented. This approach presents the advantage that the authentication is based on the signal itself rather than the recording format. Thus, it is valid for usual recording devices in police-controlled telephone intercepts. In addition, our proposal allows for the introduction of relevant information such as the recording date and time and all the relevant data (this is not always possible with classical systems). Our experimental results reveal that the speech watermarking procedure does not interfere in a significant way with the posterior forensic speaker identification.

## 1. Introduction

According to [Broeders 2001] rapid technical developments in the world of telecommunications in which speech and data are increasingly transmitted through the same communication channels may soon blunt the efficacy of traditional telephone interception as an investigative and evidential tool. The gradual shift from analogue to digital recording media and the increasingly widespread availability of digital sound processing equipment as well as its ease of operation make certain types of manipulation of audio recordings comparatively easy to perform. If done competently, such manipulation may leave no traces and may therefore well be impossible to detect. A promising development in the field of authenticity and integrity examinations of audio recordings in the analogue domain is the use of Faraday crystals as pioneered by a number of Russian scientists [Grechishkin 1996]. This potential gain is offset by the widespread availability of relatively inexpensive digital sound processing equipment. As part of the chain-of-custody process, audio recordings, like all digital data, are therefore increasingly required to be authenticated by means of checksums and hash codes or other methods to ensure their integrity. The formulation of standards for the forensic examination of audio recordings as undertaken by the AES [AES 1996] is a useful initiative, which may serve to improve standards across the whole field of forensic audio examination. However, it does not provide any technical solution.

Digital technology has provided a large amount of benefits such as higher audio fidelity than analog recordings, the possibility to make exact copies of an original recording, minimize the effect of scratches and other physical defects by means of error correction codes (such as Reed-Solomon code used on CD-Audio). However, some drawbacks have also appeared such as the possibility to modify (insert and/or delete) some portions of a given recording in a transparent way that makes this alteration process inappreciable. Forensic applications require the study of the integrity and authenticity of recordings and this problem has remained unsolved for digital recordings. We believe that the solution to this problem appears changing the philosophy: instead of studying the recording support, some relevant information must be added to the audio signal, with the following desired properties:

a) This added information must be imperceptible for a human listener.
b) It must be impossible to remove or change this added information without destroying the signal itself. Thus, it cannot be a file header or a different associated file.
c) It must be possible to determine the presence of this added information and its content in an easy and univoque way (without ambiguities).

## 1.1. *Approaches to speech authentication*

In the past few years, several approaches have been presented for the authentication of digital recordings which ought to be presented in court.

Recently it has been proposed to use a electrical network frequency (ENF) criterion for determining the authenticity of a recording [Grigoras 2006]. This approach exploits the fact that even though the electrical network frequency is fixed at 50/60 Hz, a certain amount of variation is permitted. The pattern of those variations is unique for every given time instance. Furthermore the pattern is the same over a connected electrical supply net, e.g. the European power supply network. If a recording with a device connected to the power supply net is done - besides the intended signal - it also records the ENF signal. Even a battery powered appliance used in the range of the electromagnetic field of the power supply network records the ENF signal. So if one has reference data of the ENF at a certain instance of time the ENF signal of a questioned recording can be compared to this reference.

The advantage of this method is, that the recording in question does not have to be prepared for this type of investigation. On the other hand, because the ENF information is recorded in a very narrow spectral band, which is usually not even used in human speech production, it can easily be removed by simple high-pass filtering.

Another proposal was to use the serial copy management system (SCMS) which is implemented in the digital transmission standard of professional and consumer digital audio equipment [Cooper 2005]. The SCMS flags can be used to determine whether the recording presented is an original recording or whether it is possible that a forging attempt has been made. While this approach may work if the potential forger has only access to standard consumer type equipment, it does not help in case of a well equipped audio professional. Some professional sound-cards have full control over the SCMS settings, so any SCMS setting can be pretended. In addition using high quality D/A and A/D converters, will only result in unnoticeable degradation, but would also bypass the SCMS.

The addition of meta-data to the sound signal data is also proposed. The meta-data can either be some hash code, which can be derived from the digital representation of the sound. It can also be content based information, i.e. the linear prediction coefficients of the speech signal. It has been suggested to either incorporate the meta-information attached to the signal (i.e. the wav audio format allows the embedding of meta-data in the file) [Alexander and McElveen 2006] . Alternatively the meta-information could be transmitted via an additional channel.  Those approaches introduce the problem of authentication of the meta-data since with the appropriate tools this data can be easily edited.

We believe that the short-comings of the described approaches can be overcome with watermarking technology, which will be introduced in the next section.

## 2. Watermarking

Fortunately, Forensics community is not the first one to face the problem described in section 1. Digital recording technology also permits facile data access and an increased opportunity for violation of copyright and tampering with or modification of content. This problem has been solved by means of data hiding [Bender 1996] and [Steinebach & Dittmann 2003]. Here the data of interest is directly encoded into the signal, rather than into a file header, so that it impossible or difficult to replace this information by an altered one. Another advantage is, that it is not dependent on the type of encoding of the signal.

### 2.1 Introduction

Data hiding and watermarking have received a lot of attention by different research communities during the past two decades.  To avoid confusion about terminology we will provide a brief definition of terms and introduction. An extensive introduction would go beyond the scope of this article. We can point out introductory articles, which cover both ancient history and current watermarking approaches, are [Petitcolas et al. 1999  and Hartung and Kutter 1999].

*Data hiding* is the general term for hiding information on either a channel or in a medium. Often, it is also used synonymously with the term information hiding.
The former is used for example by the military for covert communication or steganography (Greek, meaning "covered writing"). In contrast to encryption, where an unauthorized listener shouldn't be able to decode the information transmitted, the intruder shouldn't even be aware of a message sent.

_______________________________________________________________

The latter would include information embedded in an image or an audio file, which an uninformed user should not be aware of. This is also a form of watermarking.

*Watermarking* is the embedding of information in a medium, which functions as a host for this additional information. It can either be visible or invisible for the general user.

Watermarking originates in the marking of paper by mills as a means of identifying origin and quality of a sheet of paper. This is still used today sometimes for high quality paper products. A very sophisticated form of paper watermarking is used as a security feature of money bills. A form of visible digital watermarking is sometimes found in images for copyright protection, where information about the copyright status is visibly embedded in the picture (see figure 1).

In contrast invisible or transparent digital watermarks are desired not to interfere with the perception of the original multimedia content. This is the main focus of research for digital media watermarking. Among others, they are used for the purpose of copyright protection, identification, annotation and authentication of digital media contents.

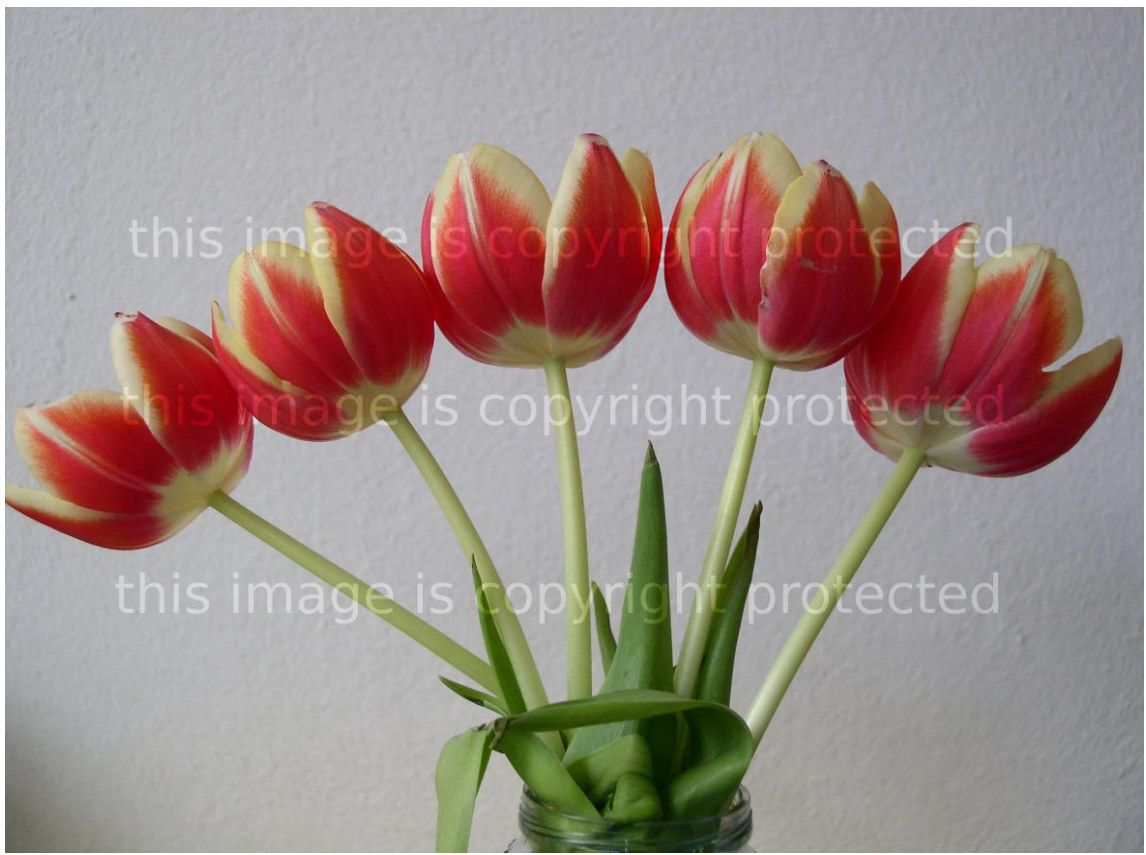

**Figure 1: Visible watermarking. The copyright information is directly embedded in the image and visibly for the viewer of the image.**

There are several, often conflicting characteristics which are of interest for watermarking.

---

- *Quantity:* Of course one wants to embed as much data as possible in a given media content. One important factor for the embedding capacity is the bandwidth of the host signal, i.e. one can be put much more additional data into a video stream, than on a telephone speech channel.
- *Robustness* describes the need for invariance of the hidden data when a "host" signal is subject to a certain amount of distortion and transformation, e.g., lossy compression, channel noise, filtering, resampling, cropping, encoding, digital-to-analog (D/A) conversion, and analog-to-digital (A/D) conversion, etc.
- *Security*: The embedded data should be immune to intentional modification attempts to remove or manipulate the embedded data. For authentication purposes this would be of high importance.
- *Fragility:* Some applications, often used for data authentication do not allow any modification of the digital media content, so it is the opposite of a robust watermark. The embedded data is lost, as soon as any modification of the signal is performed. In case of a lost watermark the authenticity of the media cannot be guaranteed anymore
- *Transparency*: This only relates to invisible watermarks, where the goal is to minimize the perceptual impact of the watermark on the host content.
  The host signal should be no objectionably degraded and the embedded data should be minimally perceptible.
- *Recovery*: The embedded data should be self-clocking or arbitrarily re-entrant. This ensures that the embedded data can be recovered when only fragments of the host signal are available, e.g., if a sound bite is extracted from an interview, data embedded in the audio segment can be recovered. This feature also facilitates automatic decoding of the hidden data, since there is no need to refer to the original host signal.

As mentioned before, some of those properties are competitive. For example, one can not both maximize security and data quantity. Depending on the application, for some characteristics a certain trade off has to be made in favor of more important properties.

There has been a lot of research effort for image watermarking and to a lesser extent for music. The main commercial application so far is copyright protection. For instance, commercial products such as Corel Draw and Photoshop include an option for "watermarking" based on Digimarc technology.

It is important to emphasize that this technique is not related to encryption, because watermarking cannot restrict or regulate access to the host signal, but rather to ensure that embedded data remain inviolate and recoverable.
In addition, the encryption process is reversible. This means that once a hacker is able to decrypt a file, this file is not protected anymore in the future. On the other hand, watermarking is an irreversible procedure, and the watermark cannot be removed without altering the quality of the host signal that much, so it is worthless for any other future usage. Watermarking can provide protection of intellectual property rights, give an indication of content manipulation, and provide a means of annotation.

_______________________________________________________________

In summary, a general watermarking system (see Fig. 2) consists of an embedding block, which overlays the host data with the additional information. The watermarked signal will be transmitted on an arbitrary channel to the receiver of the message. On this channel both inoffensive modification, such a compression and malicious attacks, which aim at modification or removal of the watermark. On the receiver side, the embedded data can be extracted again and in the case of an authentication application should give information whether the content of the media is an original or a modified version.

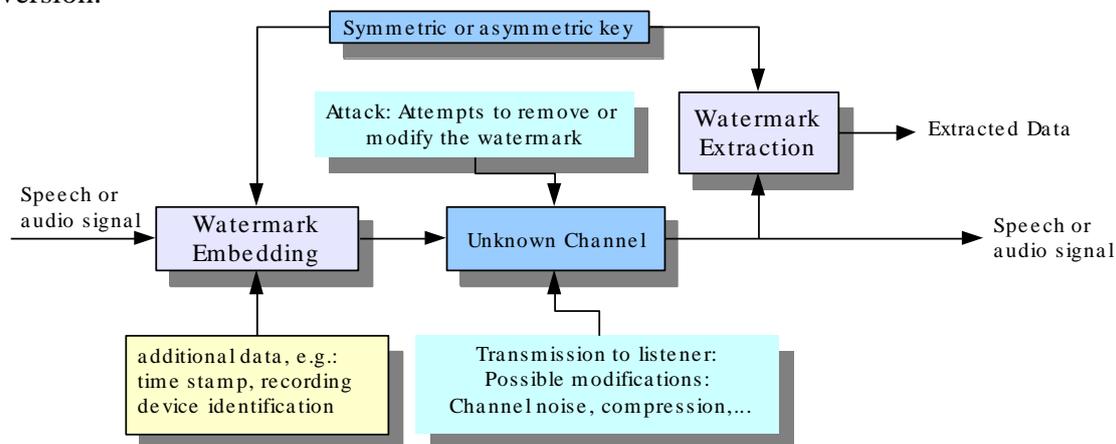

**Figure 2: Basic watermarking system. Additional data is embedded into the audio stream. Transmitted over an unknown channel and extracted at the receiver side.**

While these properties are common for all digital contents, some particularities apply depending on the media type (image, video, text, music, voice, etc.). In this paper, we will focus on speech signals. Next section describes some characteristics of watermarking of speech signals.

## 2.2 Watermarking of speech signals

Audio watermarking research has focused on copyright protection for digital audio recordings. However, the lack of commercial applications for speech has slowed down the number of publications related to this topic. Watermarking for speech signals is different in some aspects than audio watermarking due to the much narrower signal bandwidth and the special properties of a speech signal. Compared to the 44.1 kHz sampling rate for CD-audio, telephony speech is usually sampled at 8 kHz. Therefore, less information can be embedded in the signal. For perceptual hiding usually the masking levels have to be calculated. The commonly used algorithms have their origin in audio compression, such as the widespread mpeg audio coder. They are optimized for CD-audio bandwidth and computationally very expensive, since compression usually does not need to be performed in real-time. Another difference is the expected channel noise. CD-audio recordings usually have very low noise. The audio signal looses its commercial value if the background noise rises beyond a certain threshold. On the other hand, the distortion allowed for a watermark signal is very low as well. Both artist and consumer would not trade in a reduced recording quality for better copyright protection. Speech, on the other hand, is very often transmitted over noisy channels, in particular

_______________________________________________________________

true for air traffic control voice communication and telephone speech transmission. On the one hand, the channel noise is a disadvantage, on the other hand this allows much more power for the watermark signal, since the channel noise will cover it anyway. In some applications the listener expects a certain amount of noise in the signal. A summary of the differences can be seen in table 1.

|  | CD-Audio Watermarking | Speech watermarking |
|---|---|---|
| Channel noise | Very low | Can be high |
| Bandwidth | Wideband (20 kHz) | Narrowband (< 4 kHz) |
| Allowed distortion | Not perceivable | low |
| Processing delay | No issue | Very low (for real time communication) |

**Table 1: Differences between CD-audio and speech watermarking.**

In our previous works we developed a watermarking system for air-traffic control [Hagmüller 2005] (speech communication among the airplane pilot and the air traffic controller at the airport), with the goal to improve air traffic security. It is based on a spread spectrum approach. This means that low bandwidth information is modulated with a broad-band pseudorandom signal, so it is spread over all the available bandwidth. Then we proposed in [Faundez-Zanuy 2006] an enhanced biometric security system by combining watermarking with biometric speaker recognition [Faundez-Zanuy 2005]. This proposal solves the problem of replay attacks that can suffer a biometric recognizer performing over a remote channel. In this case, we used the watermark for introducing an expiry date that makes worthless the use of a pre-recording of the genuine user. The watermarking system description can be found in [Faundez-Zanuy 2006]. An advantage of our proposed scheme is that the perceptual weighting hides the watermark below the formant peaks of the speech spectrum. Thus, the "noise" of the watermark is introduced on those portions of the spectrum where the signal has a considerable amplitude. This has two important effects:

a) The perceptual quality is high for a human listener.
b) The LPC envelope is well-preserved, so the posterior LPCC parameters used by the speech recognizer will be less corrupted than their counterpart without perceptual weighting.

Figures 3, 4 and 5 illustrate this behaviour. Figure 3 shows the waveform representation of an original vowel fragment, and the same frame with and without perceptual weighting; it is clearly apparent that the weighted watermark introduces fewer artefacts. Looking at the LPC envelope (fig. 4) or the periodogram (fig. 5) we can see that the weighting procedure yields a signal closer to the original one. Thus, the LPCC parameters obtained by a recursion from the LPC coefficients [Deller 1993], will be less corrupted.

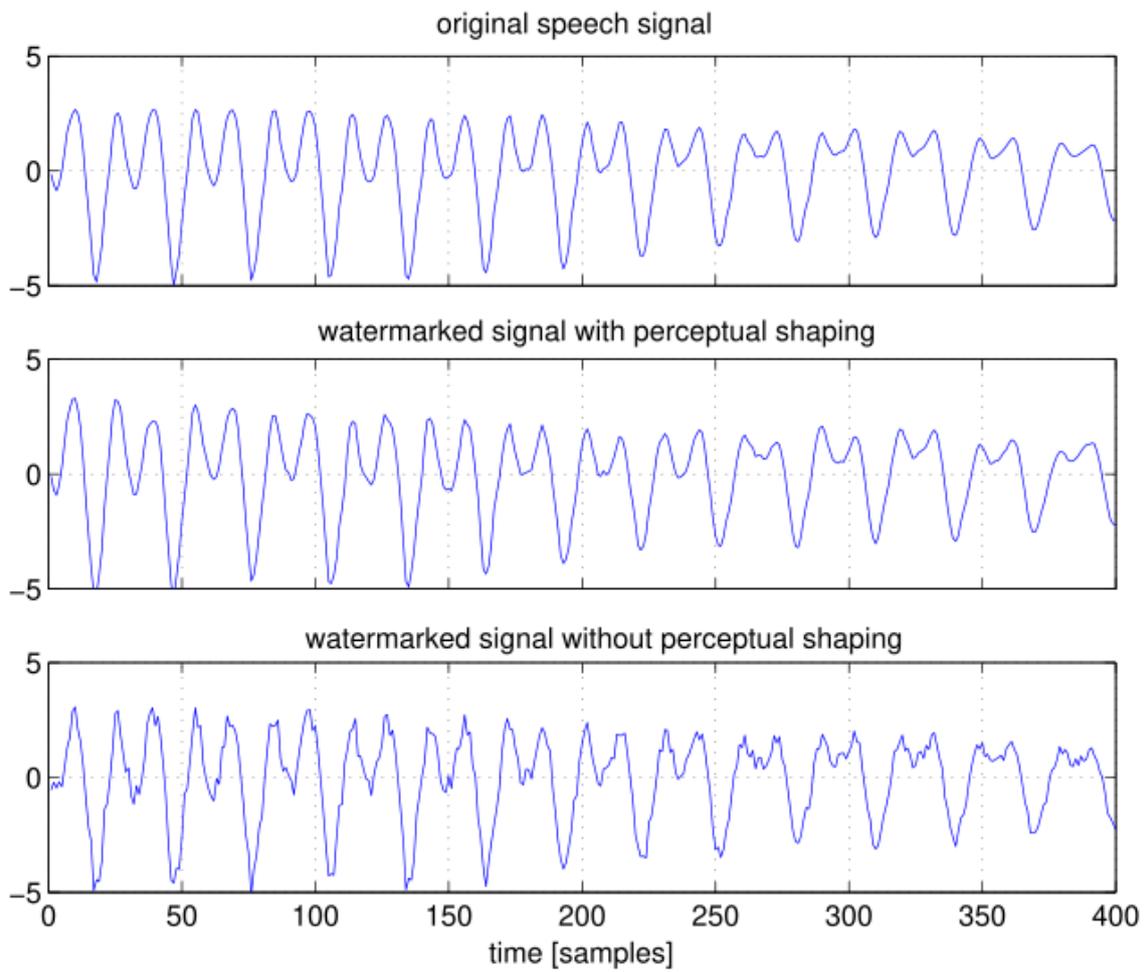

**Figure 3: Waveform plot of an example of a speech fragment, with (middle) and without (bottom) perceptual weighting compared with the original (top).**

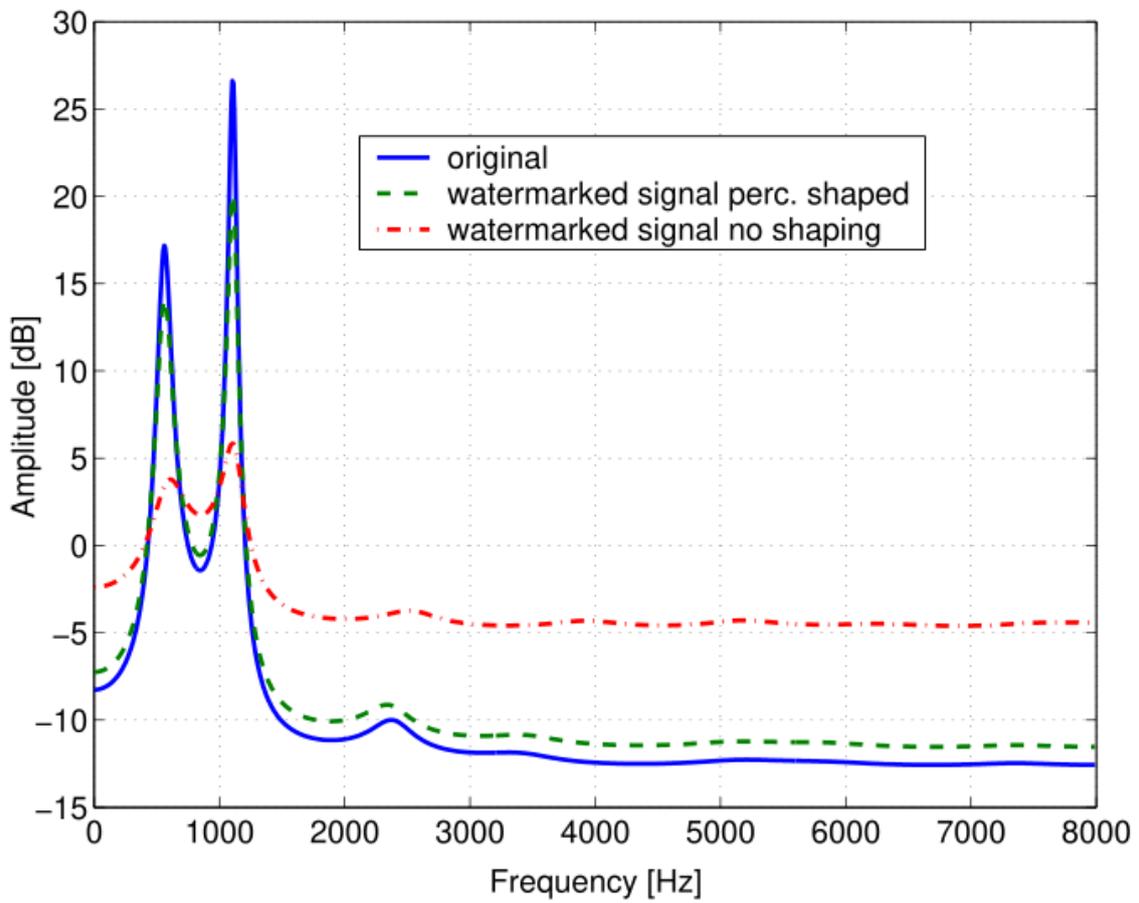

**Figure 4. Example of LPC spectrum envelope of a speech segment, with and without perceptual weighting, compared with the original.**

_______________________________________________________________________

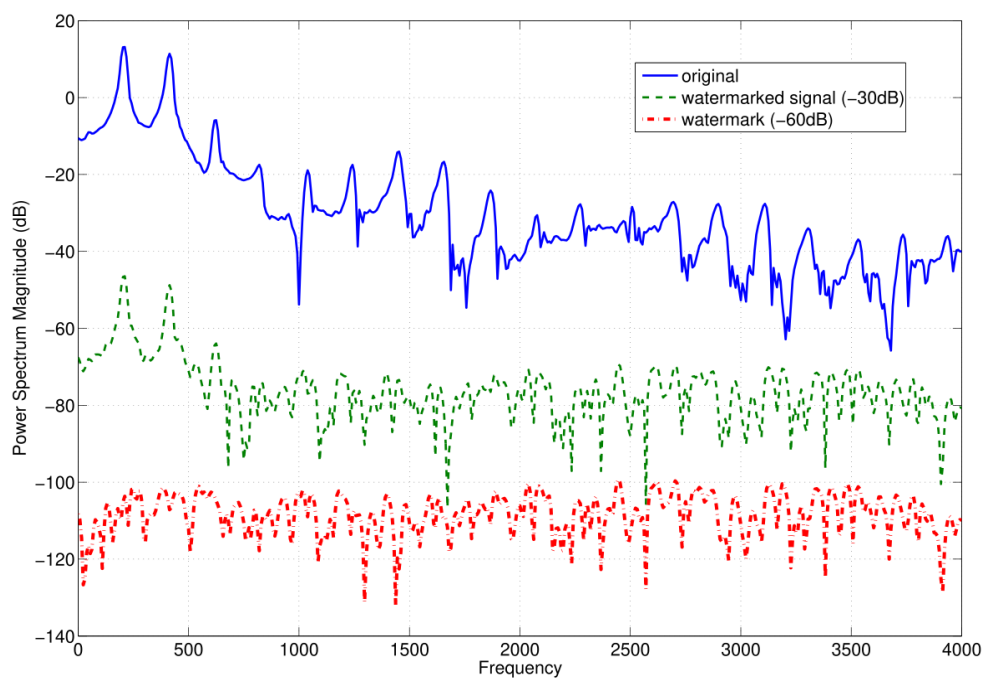

**Figure 5: Example of power spectrum density of a speech fragment, with and without perceptual weighting compared with the original.**

The good results obtained in our previous work [Faundez-Zanuy 2006] motivated our interest for evaluating the watermarking techniques in forensic applications, as discussed in the next section.

### 2.3 Watermarking of speech signals for forensic applications

Recordings made by Law Enforcement Agencies in the exercise of their legal competences in order to prevent crimes or during crime investigations are carried out nowadays using digital technology. This technology allows us to have digital audio files, and there is a wide and varied collection of equipment, interfaces, protocols, and formats at hand in the market. The available IT security systems based on that technology make the detection of any fraudulent alteration of the original content of a recording almost impossible, reaching information protection levels never seen before with analogue technology.

Public document authenticity has always been a concern for Public Administration regarding the management of official documents. Credibility or authenticity of those documents has been protected using complex physical or chemical procedures.

By applying watermarking to digital technology, i.e. inserting it into digital files, we link two kinds of information, so that a speech or image file contains additional data (watermark) such as the owner name, date, data status (public, restricted), or any other relevant information in compliance with the criteria established by the author who marks the audio files. Watermarking can guarantee that the information stored in a file faithfully matches the purpose intended by their authors: to provide data reliability by including a signal certifying its origin. For instance, data were recorded using Police equipment.

_______________________________________________________________

However, in particular, watermarking for forensic audio purposes should fulfil the following features:

- Inaudible in order to reveal its presence only to those individuals legally entitled to know it;
- Inserted in very short speech segments which could be used in any kind of situations;
- Possibility of using automatic speaker recognition systems without a relevant efficiency loss;
- Enough flexible to insert information with informative or identity value: time-stamping, anagrams, encrypted information, etc.
- Finally, the process must be irreversible to prevent the marking process from being reversed.

Due to the fact that information gathered by Law Enforcement Agencies is submitted to other legally entitled partners (Civil, Military and Judicial Authorities, etc…), when this information involves digital data it has to be protected not only by a digital signature to guarantee its integrity and origin, but also taking into account that a part of the information contained in those protected files could be used in later processes such as copies aimed at authorised listeners. It should be possible to authenticate the origin of any digital information from Law Enforcement Agencies, regardless whether it was directly provided by these Agencies or when it involves partial or full copies of it. This goal can be fulfilled by means of watermarking, which makes possible to detect any insertion or erase processes implemented in the original recording. Therefore, watermarking improves the credibility of any copy made from the original recorded information.

For the abovementioned reasons, digital signature and watermarking are two IT security systems that are very well matched for audio police recordings. Each one has its own functionality complemented by the other. We consider both useful to be applied in any kind of digital audio Police applications.

High Court Jurisprudence requires Police Forces to submit the original recordings as a requirement to determine the unquestionable credibility of their content. This is a consequence of the presumption of innocence principle which establishes that the defendant is presumed innocent until proven guilty by the prosecution. Therefore, we consider strongly advisable to take IT security measures:

- To guarantee the integrity of original recordings;
- To make possible to detect non authorised alterations; and
- To guarantee the origin of the information when segmentation of the original recording is needed and those segments are disclosed to third parties.

## 3. Proposed speech authentication system

### 3.1 Description

Our proposed system (see Fig. 6) uses watermarking technology, such es briefly introduced in section 2.2., to embed a time stamp information into the voice signal. This should be done directly in the recording device. Our system proposal embeds a watermark, which contains the exact time of the current recording every second. The embedding and reading of the time stamp can only be performed if a key, which determines the modulation sequence, is available.

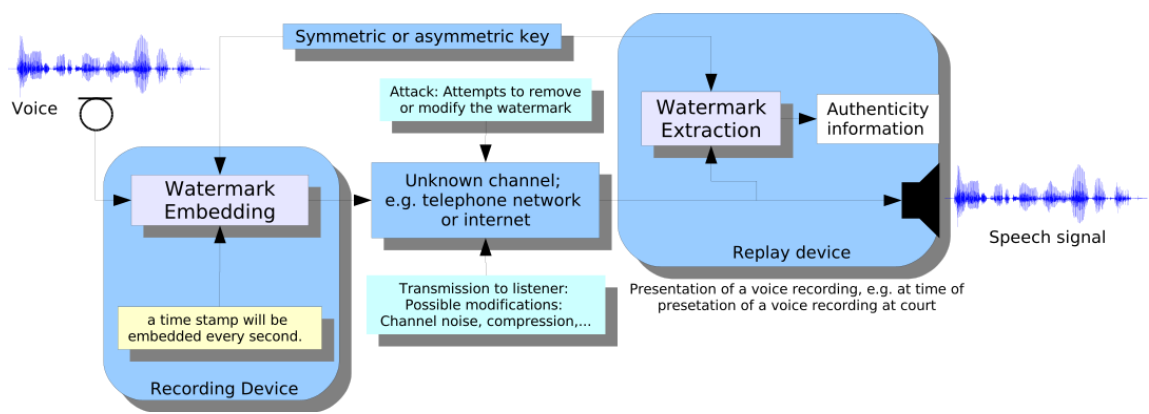

**Figure 6: Watermarking system for speech authentication.**

The watermarked signal can then be stored or distributed. In the case of e.g. the use of a voice sample in court the authenticity of the recording can be determined. If the recording has been tampered with, e.g. a word has been removed, replaced or inserted, the integrity of watermark at this time instance will be lost. Because of the high time resolution of the embedded time stamp, not only the general authenticity of the recording can be investigated, but also at which time instant a possible tampering attack has been made, which is where the continuous time stamp is broken.

Next section is devoted to experiments about the interaction between speech watermarking and speaker identification for forensic applications.

3.2 database description and experimental results

All the speech material used in the experiment belongs to the B.D.R.A. (Base de Datos de Registros Acústicos) [B.D.R.A.], a Spanish Public File belonging to the Spanish Ministry of Interior for scientific purposes. The speaker recognition system used is based on the ATVS GMM-MAP-UBM technology submitted to NIST 2004 SRE [NIST 2004].
The experiment consists of evaluating the system performance by means of a tool called "EvalIdentiVox2004" [Gonzalez-Rodriguez 2006], a script developed by AGNITIO [Agnitio] to calculate DET (detection error tradeoff) and Tippet plots using test files, speaker models, and reference populations. In this experiment 300 test files have been used, each one 10 seconds long, spontaneous speech, GSM channel, all of them from legal phone tapings carried out from 2000 to 2004 and involving voices recorded on tapes. 10 files per speaker were necessary, with a total of 30 male speakers, all of them speaking Spanish, resulting in the above mentioned 300 files. Each utterance is independent, i.e. it is absolutely different from the other ones.

2 additional minutes of speech from each one of the 30 speakers were recorded in conversations independent from the abovementioned ones. These files were used to calculate the unquestioned acoustic model for each speaker.

With regard to the reference population, it comprised 35 male Spanish speaking speakers different from the 30 earlier mentioned. The samples were obtained from legal phone tapings as well, under the same conditions.

Comparing the Tippet (see figure 7) plot in the experiment, 300 target scores were calculated along with 8.700 non-target scores. The results from the original files are depicted in blue and in red those obtained using the watermarked files. The performance differences are irrelevant. The appropriate combination between noise and speech has avoided a substantial difference in performance.

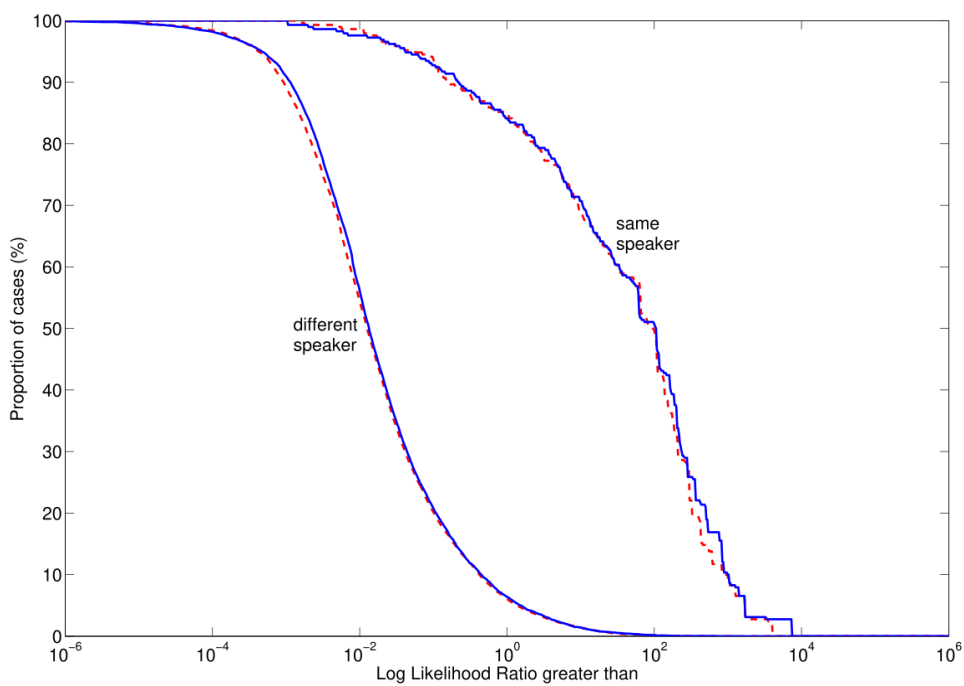

**Figure 7: TIPPET plot for a given Log Likelihood ratio. Clean (solid line) and watermarked (dashed line) signal show no significant difference in performance.**

### 4. CONCLUSIONS

In this paper we have discussed the problem of authentication of forensic audio. We have summarized the existing approaches and we have proposed a new one based on perceptual speech watermarking, which is independent of the recording device (optical, magnetic, solid state). In addition, we have performed a set of experiments on a forensic database and we have found that speech watermarking does not interfere with the posterior biometric speaker identification/verification.

Faundez-Zanuy M, Lucena-Molina JJ, Hagmüller M. Speech watermarking: an approach for the forensic analysis of digital telephonic recordings. J Forensic Sci. 2010 Jul;55(4):1080-7. doi: 10.1111/j.1556-4029.2010.01395.x. Epub 2010 Apr 16. PMID: 20412360.

_______________________________________________________________

Faundez-Zanuy M, Lucena-Molina JJ, Hagmüller M. Speech watermarking: an approach for the forensic analysis of digital telephonic recordings. J Forensic Sci. 2010 Jul;55(4):1080-7. doi: 10.1111/j.1556-4029.2010.01395.x. Epub 2010 Apr 16. PMID: 20412360.

---